\documentclass[11pt,a4paper]{article}
\usepackage{epsfig} 
\usepackage{graphicx}
\usepackage{graphpap,geometry}

\newcommand*{\e}{\mathop{\mathrm{e}}\nolimits}
\newcommand*{\ph}{\vec{\phi}} 
\renewcommand*{\d}{\mathrm{d}}
\newcommand*{\dmu}{\partial_{\mu}} \newcommand*{\umu}{\partial^{\mu}}
\newcommand*{\dnu}{\partial_{\nu}}

\title{Simulated Annealing for Topological Solitons}

\author{Mark Hale \\
{\it Centre for Particle Theory, University of Durham, Durham DH1 3LE, UK} \\
and \\
Oliver Schwindt and Tom Weidig \\
{\it Dept.~of Physics, UMIST, Manchester M60 1QD, UK}}

\begin{document}
\maketitle

\begin{abstract}
The search for solutions of field theories allowing for
topological solitons requires that we find the field configuration
with the lowest energy in a given sector of topological charge. The
standard approach is based on the numerical solution of the static
Euler-Lagrange differential equation following from the field
energy. As an alternative, we propose to use a simulated annealing
algorithm to minimize the energy functional directly. We have applied
simulated annealing to several nonlinear classical field theories: the
sine-Gordon model in one dimension, the baby Skyrme model in two
dimensions and the nuclear Skyrme model in three dimensions. We
describe in detail the implementation of the simulated annealing
algorithm, present our results and get independent confirmation of the
studies which have used standard minimization techniques.
\end{abstract}

\section{Introduction}

Solitons have played an ever increasing role in the description of
physical phenomena since their discovery by Russell \cite{russell} in
1834.  Generally speaking, a soliton is a stable localized solution of
a nonlinear partial differential equation which propagates at a
constant speed and stays localized even after an interaction with
another soliton, see Ref.~\cite{soliton:intro} for an introduction to
solitons. They are particle-like extended objects.  Well-known soliton
models in 1D are the Korteweg-de Vries (KdV) and the sine-Gordon
model.  The stability of the KdV soliton is due to the dynamical
balance between the nonlinear and the dispersive term in the KdV
equation. This differential equation belongs to the class of
integrable models which can be solved exactly. The sine-Gordon model
is also integrable and the stability of its soliton is also based on
the balance between nonlinearity and dispersion.  However, it also
belongs to the wider class of models whose solitons are stable by
conservation of a topological charge or winding number, as discussed
see in Sec.~\ref{sine}. In topology, the field is interpreted as a
mapping from physical space to field space and a field configuration
with given topological charge cannot dynamically change into a field
configuration with a different charge. In this paper, when we discuss
topological solitons, we shall consider only the field configuration
with lowest energy in a non-zero topological charge sector.

Topological solitons arise in many areas of physics: field theory
(e.g.~vortices, monopoles and instantons; as discussed in
Ref.~\cite{ryder}), condensed matter (e.g.~baby skyrmions \cite{qhe}),
nuclear physics (skyrmions, see Ref.~\cite{holzwarth:review}),
cosmology (e.g.~cosmic strings \cite{cosmic:string}) and string
theory/M-theory (e.g.~Olive-Montonen duality \cite{duff}).  They
possess many interesting properties.  The conservation of topological
charge can be used to model particle conservation and annihilation:
the number of solitons is conserved and two solitons with opposite
charge can annihilate. Since the stability of solitons is assured by
topology, there exits considerable freedom in constructing appropriate
lagrangians for physical systems.  The common constraint of Lorentz
invariance is easily imposed by using covariant terms. However, the
use of topological models has a major disadvantage. The models are
generally not integrable and few analytical techniques are available
(only Ans\"atze, topological bounds, etc.). It is therefore crucial to
study the models using reliable and efficient numerical methods.

We are looking for the (static) lowest energy field configuration in a
given topological sector. Thus, we need to minimize the energy
functional $E$ of the field theory, the integral of an energy density
$\mathcal{E}$ over a manifold $M$:
\begin{equation}
 E[C]=\int_{M}\!\d^nx~\mathcal{E}\left(x,f(x),f'(x)\right)
\label{min}
\end{equation}
where $f(x)$ is a field configuration $C$. The topology typically
imposes  some boundary conditions $f(\partial M)$. We are searching
for the function $f_{\rm{min}}(x)$ which gives the lowest value for
$E$. There are two possible  approaches: to solve the  Euler-Lagrange
equation of the functional $E$  with respect to the function $f(x)$ or
to minimize $E$ through some other means.

Hitherto only the first approach, via the Euler-Lagrange equation, has
been used with topological systems.  We shall review the standard
numerical techniques which apply shooting or relaxation methods and
discuss their reliability and ease of use.  In this paper, we show how
to minimize the energy functional directly by using the simulated
annealing (SA) algorithm, as proposed in Refs.~\cite{sc:second,tom}.
SA is based on the fact that a solid which is slowly cooled down,
assuring thermal equilibrium at each temperature, reaches its ground
state. The SA algorithm describes the cooling process and a Metropolis
subalgorithm brings a system into thermal equilibrium.  SA has been
applied to minimization problems in such diverse areas as
combinatorial optimization (such as the traveling salesman problem),
circuit design, finance, physics and military warfare: see
Ref.~\cite{ingber} and Ref.~\cite[Sec.~10.9]{num}. We give a detailed
introduction to SA and describe our implementations. We find the
topological solitons of the sine-Gordon model in 1D, the baby Skyrme
model in 2D and the nuclear Skyrme model in 3D and compare our results
to those obtained using standard minimization techniques.

\section{Minimization via Euler-Lagrange Equation}

The standard procedure uses the Euler-Lagrange equation resulting from
the variation of the functional $E$ with respect to the function
$f(x)$ to find the minimal energy solution $f_{\rm{min}}(x)$.  In the
1D case, for example, the problem is a two point boundary value
problem satisfying the differential equation
\begin{equation}
 \frac{\d}{\d x}\left(\frac{\d\mathcal{E}}{\d f'}\right)
 -\frac{\d\mathcal{E}}{\d f}=0.
\label{DE}
\end{equation}
It is a second order ODE, and a PDE in two or more dimensions, which
is equivalent to a set of first order ODEs. Let $f(a)$ and $f(b)$
represent the boundary conditions a field configuration has to satisfy
over the interval $[a,b]$. There are two standard approaches: the
shooting and the relaxation methods (as discussed in
Ref.~\cite[Chap.~17]{num}).

\subsection{The Shooting Method}

The shooting method is usually based on integration from one boundary
to the other.  The value of the function at the point $x=a$ is taken
to be $f(a)$ and an initial guess $\alpha$ for its derivative is
made. A numerical integration, for example with a Runge-Kutta method,
up to the other boundary point $x=b$ then gives an estimate
$f_{\alpha}(b)$ for $f$ at $b$.  This value is compared to the known
boundary value $f(b)$ and $\alpha$ is adjusted to match
$f_{\alpha}(b)$ closer to $f(b)$. This procedure is repeated until the
desired accuracy is achieved. The shooting method is unrivaled in
speed and accuracy, but only applicable in one dimension.

\subsection{Relaxation Methods}

Gauss-Seidel over-relaxation (SOR) is commonly used to solve the
boundary problem directly.  A time-dependent differential equation, a
diffusion equation, is constructed out of the 1D ODE (\ref{DE}),
\begin{equation}
 \frac{\d f(x,t)}{\d t}=\omega\, \d x^2 \left[ \frac{\d}{\d x}\left(\frac{\d\mathcal{E}} {\d f'}\right)-\frac{\d\mathcal{E}}{\d f} \right].
\label{omegaDE}
\end{equation}
If the system reaches equilibrium, i.e.~$\frac{\d f}{\d t}=0$, this
configuration is a solution to (\ref{DE}). One starts out with a
configuration satisfying the boundary conditions.  The coefficient
$\omega$ of the leading term, which has the form $\frac{\d^2f}{\d x^2}$,
is dimensionless and determines the speed of convergence.  The choice
of integration method is not very sensitive, we can use Euler
integration, the Runge-Kutta method or the Crank-Nicholson method.
The standard SOR uses the Euler method with updated information from
already computed field values at lattice points and ensures better
convergence.

In many cases one also studies the time-evolution of the
models. Computer codes developed for this purpose can be adapted to
find minimal energy solutions by adding a damping term to the equation
of motion.  For example in \cite{baby_pots}, two baby skyrmions are
put in an attractive configuration and form an oscillating bound
state. As the system has been made dissipative, the energy of the
system decreases with time until a minimum is reached. Thus, finding a
minimal energy solution is translated into a damped time-evolution.
The same effect can often be reached by working with a finite box and
absorbing any outward propagating radiation on the boundaries.

Relaxation techniques are well documented and applicable in any
dimension, see Ref.~\cite{num}.  The SOR method has theoretically the
best rate of convergence but might be less than optimal since the best
choice of $\omega$ can rarely be determined for a nonlinear system and
must be made by trial and error. Using damping in a time-evolution
problem is convenient, but one first has to set up the time-evolution
code. It is impossible to estimate the error on an integration step
and one needs to monitor conserved quantities. This is especially
important if, as is often the case, the field has to satisfy a
constraint. Furthermore, if the initial configuration is far from the
global minimum, we might end up in a local minimum.  Moreover, the
derivation of the corresponding Euler-Lagrange equation becomes
increasingly difficult when higher order terms are added to the
lagrangian or when complicated constraints on the field space are
present.

We have come to the conclusion that the weaker points of iterative
minimization techniques via the Euler-Lagrange equation are:
\begin{itemize}
\item Uncertainty about the global nature of the minimum obtained.
\item Lack of direct control over the integration errors (important
for constrained fields).
\item Tedious derivation for complicated lagrangians.
\end{itemize}

\section{Minimization via Simulated Annealing}

Minimizing the energy functional directly is a more straightforward
approach than solving the equations of motion and we propose to use
the flexible and easy-to-implement simulated annealing technique.

\subsection{Metropolis Principle}

In 1953 Metropolis et.~al.~\cite{me:equation} proposed an algorithm,
now called the {\it Metropolis} or {\it $M(RT)^{2}$} algorithm, that
can be used to bring a statistical system into thermal
equilibrium. The {\it $M(RT)^{2}$} is most commonly used to evaluate
thermal averages $\langle\mathcal{F}\rangle$ of a quantity $\mathcal{F}(C)$,
\begin{equation}
 \langle\mathcal{F}\rangle=\frac{\int\mathcal{F}(C)P(C)\,\d C}{\int P(C)\,\d C}.
\label{average}
\end{equation}
Here $P(C)$ is a probability distribution for configurations $C$. It
must satisfy $P(C)\!\ge\!0$ and $\int P(C)\,\d C\!<\!\infty$ in order to be
normalisable. For example, $\langle\mathcal{F}\rangle$ can stand for the thermal
average of our energy functional $E[C]$ in Eq.~(\ref{min}).  In fact,
the Metropolis algorithm is only one of the possible sampling methods
for the Monte Carlo evaluation of the integral, see
Ref.~\cite[Sec.~3.7]{ka:monte}.

For the system to reach thermal equilibrium it needs to satisfy the
condition of detailed balance,
\begin{equation}
 K(C_2\mid C_1)P(C_1)=K(C_1\mid C_2)P(C_2).
\end{equation}
Here $P(C)$ is the probability to find the system in the
configuration, or state, $C$, and $K(C_2\mid C_1)$ is the conditional
probability to move from $C_1$ to $C_2$.  The conditional probability
$K$ is usually decomposed as
\begin{equation}
 K(C_2\mid C_1)=A(C_2\mid C_1)T(C_2\mid C_1),
\end{equation}
where the transition probability $T(C_2\mid C_1)$ can be chosen to be
any normalized distribution. It is used to select a random trial move
from $C_1$ to $C_2$.  The complications are kept in $A(C_2\mid C_1)$,
which gives the probability of accepting this move and is the
correction to the arbitrarily chosen $T(C_2\mid C_1)$. The key element
of the algorithm is the evaluation of the function $A(C_2\mid C_1)$ by
a rejection technique. Thus the function $T(C_2\mid C_1)$ is sampled,
and the resulting configuration is accepted or rejected depending on
the value of $A(C_2\mid C_1)$. One usually defines
\begin{equation}
 q(C_2\mid C_1)=\frac{T(C_1\mid C_2)P(C_2)}{T(C_2\mid C_1)P(C_1)}\ge 0
\end{equation}
and
\begin{equation}
 A(C_2\mid C_1)={\rm min}(1,q(C_2\mid C_1)).
\label{accept}
\end{equation}
If the configuration $C_2$ has a lower energy than $C_1$, it is
accepted.  Otherwise, it is accepted with the probability $q(C_2\mid
C_1)$. This procedure is repeated a large number of times and
eventually the system reaches an equilibrium. Here we define an
equilibrium to be the ensemble of states where the average of the
energy does not show systematic changes.

After $L$ steps, equilibrium is established and the system fluctuates
around $\langle\mathcal{F}\rangle$. The thermal average is approximated
by the sum
\begin{equation}
 \langle\mathcal{F}\rangle=\frac{1}{N}\sum_{i=L+1}^{L+N}\mathcal{F}(C_{i}),
\end{equation}
where $C_{i}$ is a state at thermal equilibrium and $N$ is the number
of iterations over which we compute the average. We can interpret
every trial move to represent a unit of quasi-time having
passed. This can not be converted to real units of time, but it is
possible to average thermodynamic properties over quasi-time when a
system is in equilibrium. It is possible to compute the mean of the
quantity because the unit of measurement, which is the number of trial
moves, cancels.  If a trial move is rejected, the old configuration
has to be counted in any averages.

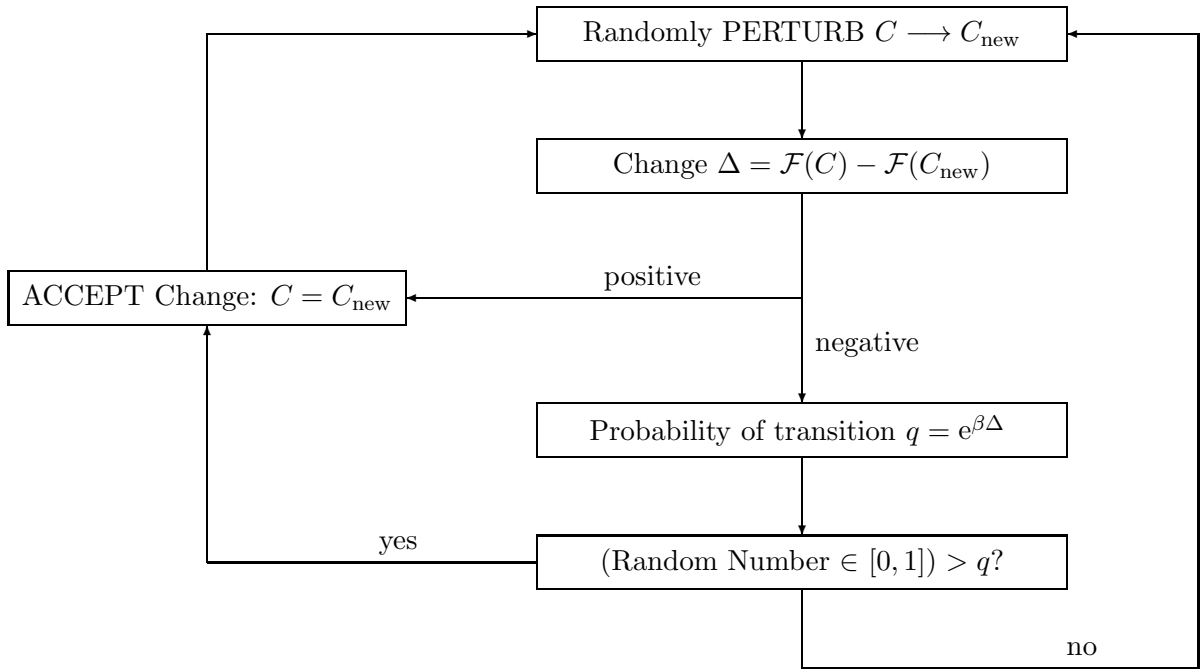
\begin{figure}
\begin{picture}(460,275)(-100,110)
\put(100,350){\framebox(200,20){Randomly PERTURB $C \longrightarrow
C_{\rm new}$}} \put(200,350){\vector(0,-1){30}}
\put(100,300){\framebox(200,20){Change $\Delta=\mathcal{F}(C)-\mathcal{F}(C_\mathrm{new})$}}
\put(200,300){\vector(0,-1){80}}
\put(205,240){negative} \put(-100,250){\framebox(150,20){ACCEPT
Change: $C=C_\mathrm{new}$}} \put(200,260){\vector(-1,0){150}}
\put(125,265){positive} \put(-25,270){\line(0,1){90}}
\put(-25,360){\vector(1,0){125}}
\put(100,200){\framebox(200,20){Probability of transition
$q=\e^{\beta\Delta}$ }} \put(200,200){\vector(0,-1){30}}
\put(100,150){\framebox(200,20){(Random Number $\in$ $[0,1])>q$?}}
\put(100,160){\line(-1,0){125}} \put(40,165){yes}
\put(-25,160){\vector(0,1){90}} \put(200,150){\line(0,-1){30}}
\put(200,120){\line(1,0){150}} \put(300,125){no}
\put(350,120){\line(0,1){240}} \put(350,360){\vector(-1,0){50}}
\end{picture}
\caption{The Metropolis algorithm: Scheme for thermal
equilibrium.}\label{metropolis}
\end{figure}

For all the examples to be discussed, $P(C)=\e^{\beta E[C]}$, where the
temperature is defined by $\beta=(k_B T)^{-1}$. If the transition
probability $T(C_{2}\mid C_{1})$ is chosen to be uniform, $q(C_{2}\mid
C_{1})=\e^{\beta\left(E[C_{2}]-E[C_{1}]\right)}$, or
\begin{equation}
 q(C\mid C_\mathrm{new})=\e^{\beta(E-E_\mathrm{new})}.
\label{prob:trans}
\end{equation}
Here, $E$ is the energy of the system in the present configuration $C$
and $E_\mathrm{new}$ is the energy of the new configuration $C_\mathrm{new}$,
that was obtained through a random change in the state of the
system (sampled from the distribution $T$). If the energy of the new
configuration $C_\mathrm{new}$ is lower, the change is accepted. If the
energy is higher, the system accepts this upward step with a
transition probability $q$.  Thus the system can escape local minima
and achieve thermal equilibrium.  In Fig.~\ref{metropolis} we show a
flow diagram representing the Metropolis algorithm. In this diagram,
$\mathcal{F}$ denotes the functional being minimized.

\subsection{Simulated Annealing}

In 1982, Kirkpatrick and others \cite{sim:science} observed a deep
analogy between annealing of solids and optimization or minimization
problems.  A solid that is sufficiently slowly cooled down, i.e.~at
thermal equilibrium at each temperature, will reach its ground state.
If the energy is the functional to be optimized or minimized, the SA
scheme should find the minimum energy function of this functional
according to a statistical proof by Geman and Geman \cite{geman}. One
starts out with a configuration at a high temperature and runs the
Metropolis algorithm. The size of change of the functional is
proportional to the temperature. Once we have reached thermal
equilibrium, the temperature is decreased according to a cooling
schedule and the procedure repeated as often as necessary.  SA is a
conceptually easy-to-understand minimization technique.

There are several varieties of SA algorithms, each designed to speed
up the minimization of a particular problem.  The application to a
minimization of a continuous problem deserves some reflection on the
discretization of the derivatives. The most important question is the
cooling schedule; a sufficiently slow cooling is crucial for the
``statistical proof of convergence''. Quite often, a slow cooling is
not needed to reach  the global minimum and a faster cooling schedule
can be used: this SA version  is called simulated quenching. For more
information on SA, we recommend reading Ref.~\cite{ingber} and
Ref.~\cite[Sec.~10.9]{num}.  There is no unique way of implementing
the SA scheme and there exits ample opportunity to improve the
code. However, every SA implementation faces the same issues.

{\bf Which initial guess?} Unlike the case of the relaxation method,
the initial configuration is not important as the system should be
able to jump out of local minima. However, an initial guess close to
the global minimum solution can lead to a reduction of the running
time.

{\bf What sampling method to use?} The changes to $C$ should be made
such that the configuration space is well sampled. As the temperature
decreases, so should the size of the changes. Usually the choices made
are random in the configuration space with a Gaussian  or Lorentzian
distribution. However, since we are dealing with functionals rather
than functions such steps are expensive to compute. Therefore, we
restrict  ourselves to changes at individual gridpoints.

{\bf At what initial temperature to start?} A high temperature puts
the system in thermal equilibrium quickly, but at too high temperature
the soliton unwinds. Too low a choice for the initial temperature can
leave the system in a local minimum. The best choice is found by trial
and error.

{\bf When is equilibrium reached?} The determination of the
equilibrium position is crucial and a statistical study on the changes
of the system is essential. Equilibrium is reached when the energy of
the system fluctuates, but does not show any systematic trend.
However, meta-stable states have been observed (like glasses), where
the state of the system changes so slowly that one runs the risk to
interpreting it as being constant.

{\bf The choice of cooling schedule.} The temperature should decrease
according to a log rule to assure convergence to the global minimum
(see Ref.~\cite{geman}).  However, it takes a long time to reach
thermal equilibrium with such a cooling schedule.  Often, the cooling
is speeded up by an exponential cooling schedule using big temperature
decreases or a weaker equilibrium condition.

{\bf Discretization of continuous functional $\mathcal{F}$?} It is
important not to use the central difference, because it does not
depend on the function at the center point. However, this problem can
be  overcome by computing the derivatives midway between two
gridpoints. We discuss this later.

{\bf Use of constraints?} Constraints are no problem, because we use
random changes that satisfy the constraints.

\section{Simulated Annealing in 1D}

We have used the sine-Gordon model for our 1D SA implementation,
because it is one of the simplest field theories exhibiting extended
structures and it is exactly solvable, see Ref.~\cite{raj}.  The
sine-Gordon model is also a very good toy model for solitonic quantum
field theories, for the quantum mass correction and the S-matrix are
exactly known. Further, Coleman \cite{coleman:sine} has shown that the
quantum sine-Gordon model and the massive Thirring model are dual to each
other:  the bosonic soliton in sine-Gordon is a fermion in the massive
Thirring model. Finally and most importantly for this paper, the
soliton solution is known exactly and we can compare it to our SA
results.

\subsection{The Sine-Gordon Model}\label{sine}

The sine-Gordon model is described by the lagrangian density
\begin{equation}
\mathcal{L}=\frac{1}{2}\dmu\phi\,\umu\phi-(1-\cos\phi).
\end{equation}
For simplicity, we have set the mass and the coupling to one.  The
lagrangian is invariant under $\phi\longrightarrow\phi+2\pi n$,
$n \in \mathbf{Z}$. Here $\phi(x,t)$ is an angle in field space, the circle
$\mathrm{S}^1$. The field has to go to the vacuum sufficiently fast for the
soliton to be localized and of finite energy. Therefore, we can
identify the spatial infinity in each direction with one single point
and compactify the one-dimensional space $\mathbf{R}^1$ to $\mathrm{S}^1$.
The field theory of the sine-Gordon model can be described by the map
\begin{equation}
 \phi(t): \mathrm{S}^1 \longrightarrow \mathrm{S}^1
\end{equation}
at a given time $t$. This non-trivial mapping gives us the possibility
to partition the space of all possible field configurations into
equivalence classes having the same topological charge or winding
number. We can visualize this concept with a belt. We can trivially
close it or we can twist one side by 180 degrees and close it or we
can anti-twist it by 180 degrees, i.e.~twist it by $-$180 degrees, and
close it. The twist in the belt cannot be undone unless one opens the
belt. Topological solitons can be thought of as twisted field
configurations. The homotopy group $\Pi_1(\mathrm{S}^1)=\mathbf{Z}$ describes the
twists in the map.  For example, if we twist the belt twice and then
anti-twist it twice, we get back to an untwisted belt: very much like
an annihilation process in particle physics.  The `twist', i.e.~the
topological charge, is fixed by boundary conditions and conserved.

The corresponding Euler-Lagrange equation for the sine-Gordon model is
\begin{equation}
\ddot{\phi}-\phi''+\sin\phi=0.
\end{equation}
One can find the minimal energy solution by solving the static version
using theoretical or numerical methods. The 1-soliton, i.e.~minimal
energy solution of topological charge one, can be derived from the
Bogomolnyi equation (see Ref.~\cite[Sec.~2.5]{raj}):
\begin{equation}
\phi'=\pm \sqrt{2(1-\cos\phi)}.
\end{equation}
Rewriting this in terms of $\sin(\phi/2)$, integrating and inverting
the resulting relation, one finds
\begin{equation}
\phi_{\rm st}(x)=4\arctan [\exp\left(x+x_0\right)].
\label{sine:soliton}
\end{equation}
We can derive the solitons with higher charge via a Backlund
transformation \cite{raj}.  The static minimal energy solution
satisfies the boundary condition $\phi(-\infty)=0$ and
$\phi(\infty)=2\pi$, the field winds around the field sphere
$\mathrm{S}^1$ once. The expression of the energy density is
\begin{equation}
\mathcal{E}(x)=4\sin^2\left[\phi_{\rm{st}}(x)\right].
\end{equation}
The energy goes to zero at spatial infinity and the integral is
finite.  The total energy is $\int\!\mathcal{E}(x)~\d x=8$. In the next
section, we discuss the 1-soliton, Eq.~(\ref{sine:soliton}), and
calculate its total energy using the SA scheme.

\subsection{Implementation of Simulated Annealing}

Three aspects of the SA implementation  are crucial for successful
minimization: the derivatives, the sampling method and the cooling
schedule with thermal equilibrium.

\subsubsection*{Derivatives}

The most accurate discretized derivative is the centered difference.
However, this causes problems with derivative terms as it does not
depend on the function at the point where the energy is being
evaluated.  This results in a decoupling between neighboring points
which gives rise to two independent sub-lattices.  The configuration
becomes spiky since the values jump between the two sub-lattices.  To
avoid this problem, the energy is computed between the gridpoints
rather than at the gridpoints.  The value of the function between the
gridpoints is taken to be the average of the values at the surrounding
points.
\begin{eqnarray}
\phi(x_{i+\frac{1}{2}}) & = & \frac{\phi(x_i)+\phi(x_{i+1})}{2}, \\
\label{central}
\frac{\partial\phi(x_{i+\frac{1}{2}})}{\partial x}
& = & \frac{\phi(x_{i+1})-\phi(x_{i})}{\d x}.
\end{eqnarray}

\subsubsection*{Sampling}

Typically, SA is used to minimize a function $f(x)$ with $x$ being a
vector.  The general form of a change to a configuration is
\[ x_i\rightarrow x_i+M_{ij}U_j \]
where $M_{ij}$ is a matrix, and $U_i$ is a vector of random numbers
satisfying an appropriate probability distribution, see
Ref.~\cite{sa:cont}.  The matrix $M_{ij}$ needs to be chosen such that
the configuration space is well-sampled.  Information from the cooling
process can be used to dynamically adjust $M_{ij}$.  We are interested
in minimizing energy functionals on a lattice of $N$ gridpoints so our
$x$ vector will have $N$ components.  This makes calculating a new
configuration quite an intensive process.  To simplify matters we
sweep across the grid changing individual points at a time.  The
random numbers $U_i$ are taken from a Lorentzian distribution, rather
than a Gaussian distribution.  This is a quite common modification to
the original SA algorithm as the Lorentzian has a longer tail.  The
mean and width of the distribution need to be chosen so that we get a
good sampling of configuration space.  A narrow distribution will only
sample the local neighborhood while a wide distribution will spend too
much time probing irrelevant configurations.  To achieve a good
balance the width is adjusted so that 50\% of all the proposed new
configurations are accepted (this is called the acceptance rate).  If
the mean is taken to be linearly dependent on the temperature then the
acceptance rate will remain roughly constant throughout the cooling.
This leaves the constant of proportionality to be determined at the
start of the cooling process.

\subsubsection*{Cooling Schedule and Thermal Equilibrium}

We use an exponential cooling schedule; the temperature is decreased
by a fixed ratio at each cooling step.  This violates Geman and
Geman's statistical guarantee of reaching the minimum solution.  Since
we do not expect many local minima in 1D this should not be a problem
here.

There are several approaches to determining whether the configuration
is in equilibrium.  A popular one is based on a sliding average, also
known as binning, where the mean energy calculated over a number of
iterations is monitored to see whether it has converged to a fixed
value.  We employ a simpler, related condition which monitors the
lowest energy obtained during a set sequence or chain of iterations
until no new low from one chain to the other is found.  Since
equilibrium is by its very nature statistical it is important to know
how many iterations need to be sampled.  A ballpark figure seems to be
10 to 15 samples per point on average.  We chose the total number of
points in the chain to be 10$N$, where $N$ is the number of gridpoints
and checked empirically that this was enough.

\subsection{Results of Simulated Annealing}

\begin{figure}
\begin{center}
\input{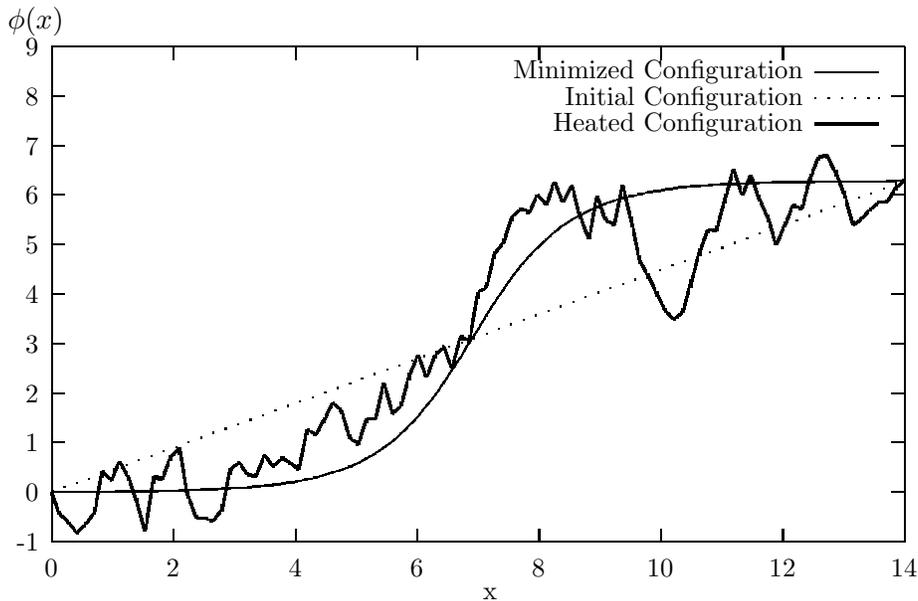}
\end{center}
\caption{1D SA cooling for the sine-Gordon model.}
\label{cooling1D}
\end{figure}

\begin{figure}
\begin{center}
\input{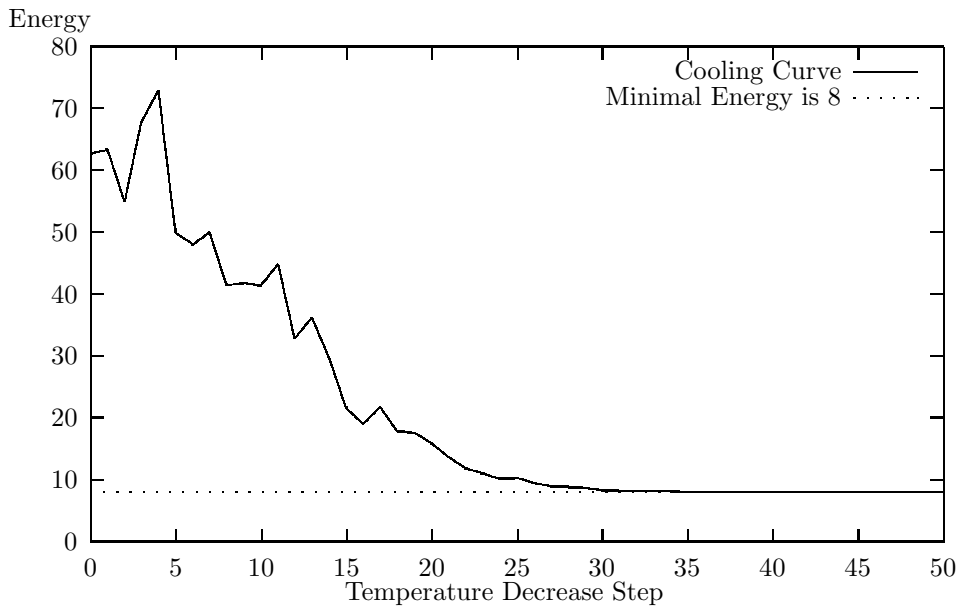}
\end{center}
\caption{Typical cooling curve for 1D SA in the sine-Gordon model.}
\label{coolingcurve}
\end{figure}

Specifically, we look here at the sine-Gordon model where we need to
minimize the following energy functional
\begin{equation}
E[\phi]=\int_{x=-a}^{x=a}\!\d x\;
\left(\frac{1}{2}\partial_i\phi\,\partial_i\phi+1-\cos\phi\right).
\end{equation}
We impose a winding or topological charge of one by setting
$\phi(-a)=0$ and $\phi(a)=2\pi$. We could use a 2D constrained field
to represent $\mathrm{S}^1$, i.e.~$\ph=(\phi_1,\phi_2)$ with
$\ph\cdot\ph=1$. However, we opted for an angle representation,
because it allows us to use fixed boundaries and the soliton cannot
unwind. In the 2D constrained coordinates, the winding is a twist in
the configuration over the whole grid and a very big fluctuation
induced by a high temperature undoes the twist.

We use different grid sizes. In Figs.~\ref{cooling1D} and
\ref{coolingcurve}, we represent different aspects of the cooling of a
sine-Gordon soliton. We start out with an initial field configuration,
here a straight line satisfying the boundary conditions. We then heat
up the configuration until thermal equilibrium is reached (thick solid
line in Fig.~\ref{cooling1D}). We cool it down by slowing decreasing
the temperature after reaching thermal equilibrium. Finally, we obtain
a minimal energy solution close to Eq.~(\ref{sine:soliton}).

We set the acceptance rate to $50\%$ and take $10N$ for the length of
the monitored chain to test thermal equilibrium. A change in these
values effects the speed and quality of minimization. We have re-run
the minimization under the same setting and find the same energy
value. This is a good indicator that the monitored chain is long
enough to obtain thermal equilibrium.  Our box size is 20 units. The
more points we use the closer the result becomes to the exact soliton
energy, namely eight, see Table~\ref{1dtable}.

To conclude, we find our SA code to be a very convenient minimization
technique in 1D. We have successfully tested our 1D SA code on many
different models.  The implementation of the SA search for solitons is
faster, for we did not have to derive the  Euler-Lagrange equation.

\begin{table}
\begin{center}
\begin{tabular}{|c||c|c|} \hline
Points & Energy \\ \hline
51 &  8.035 \\
101 & 8.009 \\
201 & 8.002 \\
301 & 8.001 \\ \hline
\end{tabular}
\end{center}
\caption{Energy versus number of points used for the sine-Gordon model.}
\label{1dtable}
\end{table}

\section{Simulated Annealing in 2D}

We have used the baby Skyrme model \cite{baby_pots} for our 2D SA
implementation, because there are exact and numerical solutions
available which we can compare to our SA results. The baby Skyrme
model is used to study some aspects of the quantum Hall effect (see
Ref.~\cite{qhe}) and is a convenient (2+1)D toy model for the (3+1)D
nuclear Skyrme model (see Ref.~\cite{holzwarth:review}) which requires
much more computational resources.

\subsection{The Baby Skyrme Model}

The nonlinear $\sigma$-model is described by the lagrangian
\begin{equation}
\mathcal{L}=\frac{1}{2}\dmu\ph\cdot\umu\ph
\label{sigmamodel}
\end{equation}
where $\ph$ is a three-dimensional field vector on the sphere $\mathrm{S}^2$
i.e.~$\ph\cdot\ph=1$. The field at a time $t$ is a map
\begin{equation}
 \phi(t): \mathrm{S}^2 \longrightarrow \mathrm{S}^2
\end{equation}
and the associated homotopy group is $\Pi_2(\mathrm{S}^2)=\mathbf{Z}$. The
existence of the topological charge, i.e.~the twisted field
configuration representing a soliton, is ensured by topology. It is
given by
\begin{equation}
B=\frac{1}{8\pi}\int\!\d^2x~{\epsilon}^{\mu\nu}\ph\cdot(\dmu\ph\times\dnu\ph).
\end{equation}

However,
we also need to make sure that the soliton has a stable scale.  From
Derrick's theorem \cite{derrick}, the energy functional corresponding
to Eq.~(\ref{sigmamodel}) is scale invariant.  A change of scale does
not change the energy and therefore numerical errors can significantly
change the scale of the soliton. It is therefore necessary to add
extra terms to stabilize the soliton, fixing the scale.  If we were to
extend the model to (3+1)D, the $\sigma$-model term would lead to an
expanding soliton and a balancing term needs to be added to ensure
stability.

The baby Skyrme model is a modified version of the $\mathrm{S}^2$
$\sigma$-model and the lagrangian is
\begin{equation}
\mathcal{L}=\frac{1}{2}\partial_{\mu}\ph\cdot\partial^{\mu}\ph
-\theta_{S}\left[(\partial_{\mu}\vec{\phi} \cdot
\partial^{\mu}\vec{\phi})^{2} -(\partial_{\mu}\vec{\phi} \cdot
\partial_{\nu}\vec{\phi}) (\partial^{\mu}\vec{\phi} \cdot
\partial^{\nu}\vec{\phi})\right] - \theta_V V(\ph).
\end{equation}
The addition of a potential and a Skyrme term to the lagrangian
ensures stable solitonic solutions. The Skyrme term has its origin in
the nuclear Skyrme model and the baby Skyrme model can therefore be
viewed as its (2+1)D analogue.  Further, in (2+1)D, a potential term
is necessary in the baby Skyrme models to ensure stability of
skyrmions; this term is optional in the (3+1)D nuclear Skyrme model.
One drawback of the model is that the potential term is free for us to
choose. The most common choices are: $V=(1+\phi_3)^4$ (the holomorphic
model has an exact 1-skyrmion solution, see
Ref.~\cite{pz:93}), $V=(1+\phi_3)$ (a 1-vacuum potential
studied in Ref.~\cite{psz:95a}) and $V=(1-\phi_3)(1+\phi_3)$ (a
2-vacua potential studied in Ref.~\cite{baby_pots}). Except for the
first choice, no closed form minimal energy solutions are known. The
baby Skyrme model is a non-integrable system and explicit solutions to
its resulting differential equations are nearly impossible to
find. Numerical methods are the only way forward.

\subsection{Implementation}\label{ThePartitionFunction}

Our 2D and 3D SA implementations originate from a more general
framework, the study of phase transitions in topological systems at
finite temperature and densities \cite{os:finite}.  The thermodynamic
partition function describes a system at a given temperature and can
only be evaluated numerically in the Skyrme models. The evaluation of
thermal averages, as discussed before, can be done with Monte Carlo
techniques and the Metropolis principle is one of the possible
sampling techniques. Conveniently, the thermal average of the energy at
zero temperature is equivalent to the minimal energy of the energy
functional to be minimized.

We start out with the grand canonical partition function
\cite{ma:statistical}
\begin{equation}
\mathcal{Z}(\beta,V,\mu) =\int_{\rm all~x_1,\dots,x_N}
\int_{\rm all~p_1,\dots,p_N}
\d^nx_1 \, \d^np_1 \cdots \d^nx_N \, \d^np_N \sum_{i=0}^{N}\exp[\beta(\mu_i-E_i)],
\label{part}
\end{equation}
where $E_i$ is the energy of the $i$th particle system at temperature
$\beta=(k_B T)^{-1}$ and $V$ is the integration range.  The integral
ranges over all phase space and is $2nN$ dimensional, where $n$ is the
number of space dimensions.  The thermodynamic partition function for
the baby Skyrme model has the form
\begin{equation}
\mathcal{Z}(\beta,V,\mu)=\int\prod_{k}\delta(\ph_{k}\cdot\ph_{k}-1)\,\d^{3}\phi_{k}
\left(\sqrt{\frac{1+\partial_{i}\ph_{k}\cdot\partial_{i}\ph_{k}}{\det(\mathcal{M}_{k})}}\right)\exp[-\beta(\mathcal{V}_{k}-\mu\mathcal{B}_{k})].
\label{BabyThermoPart}
\end{equation}
Here $\mathcal{M}$ is the mass density matrix, $\mathcal{V}$ the potential
energy density, and $\mathcal{B}$ is the topological charge density. The input
parameters of the thermodynamic partition function are the temperature
$\beta$, the volume of the system $V$ and the chemical potential $\mu$.
The $\delta$-function is required due to the constraint on the $\ph$ field.

At zero temperature, the factor in front of the exponent becomes
irrelevant and if we further set $\mu=0$,  the thermodynamic partition
function reduces to  the integral
\begin{equation}
 Z=\int\prod_{k}\delta(\ph_{k}\cdot\ph_{k}-1) \, \d^{3}\phi_{k}\exp(-\beta\mathcal{V}_{k}),
\label{Zfunction}
\end{equation}
where $\mathcal{V}_k$ is the potential energy density.  The
implementation of $\mathcal{Z}$ is similar to the implementing of $Z$,
but $\mathcal{Z}$ contains information that is not necessary for finding
minimal energy solutions The value of $Z$ is not of interest to us,
but the $\ph$ distribution as $\beta\rightarrow\infty$ is.  The
probability density function which is sampled at every point $k$ when
applying Monte Carlo is the sum over all neighbors of $k$,
\begin{equation}
P_{k}=\exp\left(-\beta\sum^{\stackrel{\rm
{number~of}}{\rm{neighbors}}}_{i=1} \mathcal{V}_{i}\right).
\label{pfunction}
\end{equation}
Examples of Monte Carlo calculations using a grand canonical ensemble
are given in Refs.~\cite{bi:monte,fr:understanding}. A good discussion
of possible errors and how to deal with them is given in
Ref.~\cite{fr:understanding}.

\subsubsection*{Monte Carlo for the Baby Skyrme Model}
\label{MCBaby}\label{MCBabyImportance}

We apply the Metropolis principle in the simplest possible way and
select a new vector $\ph_{\rm new}(x_k)$ at a gridpoint $k$ from a uniform
probability distribution function over the unit sphere, as the
integration measure with the $\delta$-function implies. We choose each
of the components $\phi_{a}$ uniformly between $-1$ and $1$. If
the sum of the squares of the components
$\phi_{1}{}^{2}+\phi_{2}{}^{2}+\phi_{3}{}^{2}$ is larger than one the
sample is rejected. All accepted vectors are scaled to obtain unit
length \cite{ka:monte}. The transition probability of this simple
method is
\begin{equation}
 T(C_2\mid C_1)=\left\{\begin{array}{cl}
\frac{1}{\rm surface~area~of~sphere} & {\rm on~the~unit~sphere} \\
0 & {\rm otherwise}
\end{array}\right.,
\end{equation}
and therefore
\begin{equation}
 q(C_2\mid C_1)=\frac{P(C_2)}{P(C_1)},
\end{equation}
where the present vector $\ph_{\rm present}(x_k)\in C_1$ and the
newly selected $\ph_{\rm new}(x_k)\in C_2$. The quantity $q$ is the
new integrand of the partition function (\ref{pfunction}) divided by
the present one. It is easiest to test a trial move for one gridpoint
at a time, although other methods will be discussed. The acceptance
probability defined by (\ref{accept}) is calculated by
\begin{equation}
 A(C_2\mid C_1)=\min\left(1,\exp\left[-\beta\left(\mathcal{V}_{\rm new}-\mathcal{V}_{\rm present}\right)\right]\right).
\label{Zaccept}
\end{equation}
A change to the vector $\ph$ at lattice point $k$ modifies the
potential energy on the gridpoint $k$ and its neighbors only (using a
linear approximation for the derivatives). This is all the information
needed to apply the Metropolis method.  The quantities of interest to
measure are the potential energy $E$ and the topological charge $B$
which should be conserved and is a check on the numerics.

A uniform sampling of the distribution has an extremely low acceptance rate;
too many vectors are rejected. We therefore use a biased sampling
technique where a new vector $\ph_{\rm new}$ is sampled near $\ph_{\rm
present}$, the vector that it is supposed to replace. We sample
vectors in an intrinsic frame where the $z$-axis corresponds to the
present vector.  The vector
\begin{equation}
 \vec{n}^{\rm int} =(n_{1}^{\rm int},n_{2}^{\rm int},n_{3}^{\rm int})
=(\sin\theta\cos\phi,\sin\theta\sin\phi,\cos\theta),
\label{vecn}
\end{equation}
gives the components of the new vector in cartesian coordinates in the
intrinsic frame. The Euler angles that define the rotation from the
intrinsic frame to the laboratory frame are given by
\begin{equation}
 \vec{k}^{\rm lab}=(\cos\beta\sin\alpha,\sin\beta\sin\alpha,\cos\alpha).
\end{equation}

The angles $\alpha$ and $\beta$ are used to rotate the new vector
$\vec{n}^{\rm int}$ to the laboratory frame,
\begin{equation}
 \vec{n}^{\rm lab}=\left(\begin{array}{c}
n_{1}^{\rm lab} \\
n_{2}^{\rm lab} \\
n_{3}^{\rm lab}
\end{array}\right)=
\left(\begin{array}{ccc}
\cos\alpha\cos\beta & -\sin\beta & \sin\alpha\cos\beta \\
\cos\alpha\sin\beta & \cos\beta & \sin\alpha\sin\beta \\
-\sin\alpha & 0 & \cos\alpha
\end{array}\right)
\left(\begin{array}{c}
n_{1}^{\rm int} \\
n_{2}^{\rm int} \\
n_{3}^{\rm int}
\end{array}\right).
\label{BabyRotationMatrix}
\end{equation}
In the previous terminology, the new unit field vector at the
gridpoint $k$ is $\ph_{\rm new}(x_k)=\vec{n}^{\rm lab}$, and it is a
vector selected from a particular probability distribution function
which is rotationally symmetric about the present vector $\ph_{\rm
present}(x_k)~(=\vec{k}^{\rm lab})$. The vectors $\ph_{\rm new}(x_k)$
and $\ph_{\rm present}(x_k)$ are inserted into the acceptance
probability. The angle $\theta$ is sampled uniformly on $[0,A)$, where
$A\le\pi$, and $\phi$ is sampled uniformly on $[0,2\pi)$.  The
corresponding transition probability is
\begin{equation}
T(C_2\mid C_1)=\left\{\begin{array}{cl}
\frac{1}{2\pi A} & {\rm if}~1-\cos\theta<A \\
0 & {\rm otherwise}
\end{array}\right..
\label{trans}
\end{equation}
No importance sampling is imposed and the probability distribution
function is uniform, so that the acceptance rate (\ref{Zaccept}) can
be applied directly. The optimal value of $A$, which we are free to
choose, depends on the particular configuration, especially on the
topological charge and the temperature $\beta$. No $A$-dependency was
discovered in any ensemble averages other than the acceptance
rate. Therefore, we believe that this method is reliable and
efficient. We allow $A$ to vary automatically to achieve an acceptance
rate near 40\%. The choice of $A$ influences the rate at which
equilibrium is reached, which is defined as the absence of change in
the average energy over a large number of steps.  At each temperature,
the system was required to reach equilibrium before being cooled
further. This ensures that cooling does not occur too quickly.

The choice (\ref{trans}) only samples a portion, rather than the whole
of the unit sphere. This is a valid method because we are modeling a
continuous system, and therefore the vectors can reach any region in a
number of steps. As $A$ is varied automatically, the whole unit sphere
is sampled for high temperatures.  If the region which is unsampled
for low temperatures were sampled, the vectors selected there would
have virtually zero probability of being accepted. For generality, we
discuss a more rigorous method using importance sampling in the
appendix. There, new vectors are chosen from a gaussian (or other)
distribution centered around the present vector. Importance sampling
allows vectors from all over the unit sphere to be selected at any
temperature, and this might be necessary for some systems, especially
when calculating thermal averages.  The disadvantage of importance
sampling over restricting the transition probability is the increased
amount of computing time.

\subsubsection*{Calculation of Field Derivatives for a Field on $\mathbf{S^2}$}

\begin{figure}
\begin{center}
\includegraphics[width=3cm]{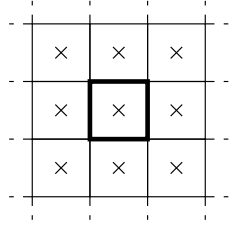}
\caption{A picture of the plaquette, where the fields are evaluated at
the intersections of the lines and the measured quantities calculated
at the midpoints $\times$.}
\label{Plaqpic}
\end{center}
\end{figure}

We calculate derivatives in a similar way to the 1D implementation
where measurable quantities are calculated in the center of the
plaquettes.  An illustration of a plaquette is given in
Fig.~\ref{Plaqpic}, where the field vectors are evaluated at the
intersections of the lines and all measurable quantities are
calculated at the midpoints.
If a field vector is altered, using the Monte Carlo method as
described above, the effect of a change has to be calculated on the
four surrounding midpoints.

All field vectors at each of the four gridpoints lie on a unit sphere
and the average of the four field vectors must also be of unit length,
because the topology requires unitarity everywhere.  A simple average
fails this criterion unless all vectors point in identical directions.
We still use the average, corrected by scaling it to unit length.

\begin{figure}
\begin{center}
\includegraphics[width=10cm]{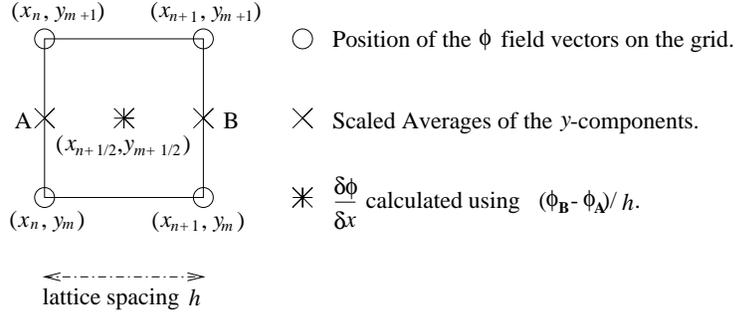}
\caption{Illustration of the scaling of the center derivative.}
\label{DerivPos}
\end{center}
\end{figure}

This also impacts the calculation of derivatives. For example, the
error in taking the $x$-derivatives is minimized if
$\frac{1}{2}\sum_{b=\{b_1,b_2\}}\phi_b (x_{n+1})$ and
$\frac{1}{2}\sum_{a=\{a_1,a_2\}}\phi_a (x_{n})$ are scaled to unit
length before the latter is subtracted from the former to obtain the
$x$-derivative. Here $\{a_1,a_2\}$ are the 2 gridpoints with the
coordinates $(x_{n},y_{m})$ and $(x_{n},y_{m+1})$, and $\{b_1,b_2\}$
are the 2 gridpoints with the coordinates $(x_{n+1},y_{m})$ and
$(x_{n+1},y_{m+1})$. These vectors are on the corners of the
plaquette, see Fig.~\ref{DerivPos}. Thus, the derivative is calculated
by
\begin{equation}
 \left.\frac{\partial\phi}{\partial x}\right|_{x_{n+\frac{1}{2}}}={\rm
 Scaled}\left[\frac{1}{2}\sum_{b=\{b_1,b_2\}}\phi_b
 (x_{n+1})\right]-{\rm
 Scaled}\left[\frac{1}{2}\sum_{a=\{a_1,a_2\}}\phi_a
 (x_{n})\right].
\label{deriv}
\end{equation}
This derivative works very well in practice. At very high temperatures
the numerics may break down, because the derivative (\ref{deriv}) is
by definition an underestimate. The 3D analog of this formula is
obtained by replacing `$\frac{1}{2}$' by `$\frac{1}{4}$' and summing
over the four components of $a$ and the four components of $b$.

\subsubsection*{Updating Mechanisms}

\begin{figure}
\begin{center}
\includegraphics[width=2.5cm]{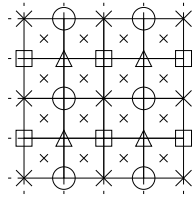}
\caption{The four subgrids for single point changes, each labelled by
a symbol.}
\label{SubgridN}
\end{center}
\end{figure}

In our 1D simulations, we have randomly selected which gridpoint
should be sampled.  In our 2D and 3D implementation, we sweep over the
gridpoints sequentially. The new vectors are stored and the changes to
the field are only updated after a complete sweep over the entire grid
to avoid unwanted sequential correlations. We have split the grid into
four independent subgrids, each labelled by a different symbol in
Fig.~\ref{SubgridN}.
The subgrids are chosen at random and, at each sweep, only one of the
possible four composing vectors of the derivatives and field averages
at the midpoints are changed. This avoids the creation of fluctuations
between neighboring vectors for high acceptance rates, which produce
an unphysical increase in energy.  Unfortunately, the change of single
gridpoints at a time does not favor collective motion, where a
localized energy distribution moves in one direction. Changing regions
of gridpoints at a time has proven to be more efficient.

\begin{figure}
\begin{center}
\includegraphics[width=2.55cm]{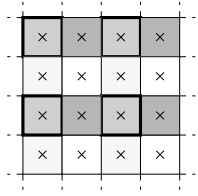}
\caption{The four subgrids for plaquette changes, each shaded
differently.}
\label{SubgridP}
\end{center}
\end{figure}

We have successfully used a plaquette updating mechanism. For a given
plaquette, we sample a vector in the intrinsic frame, see
Eq.~(\ref{vecn}). Then this frame is rotated as in Sec.~\ref{MCBaby}
for each of the four vectors on the plaquette, separately.
Fig.~\ref{Plaqpic} illustrates a plaquette surrounded by the affected
nine midpoints.  The total change in the energy density of all these
midpoints is now calculated and all four vectors are accepted or they
are all rejected. Again, we have split the grid into four subgrids,
each shaded differently in Fig.~\ref{SubgridP}. However, some
midpoints are affected by two or four plaquettes from the same
subgrid.  Therefore, we choose the subgrids randomly rather than
sequentially to avoid unwanted correlations.

\subsection{Results and Comparison}

We have investigated the three different baby Skyrme models mentioned
before. To that end, we need to minimize the energy functional:
\begin{equation}
E\left[\ph\right]=\int\!\d^2x
\left[\frac{1}{2}\partial_{i}\ph\cdot\partial_{i}\ph
+\theta_{S}\left[(\partial_{i}\vec{\phi} \cdot
\partial_{i}\vec{\phi})^{2} -(\partial_{i}\vec{\phi} \cdot
\partial_{j}\vec{\phi}) (\partial_{i}\vec{\phi} \cdot
\partial_{j}\vec{\phi})\right] + \theta_V V(\ph)\right].
\end{equation}

First, we have looked at the simplest holomorphic model with the
potential $V=(1+\phi_3)^4$. There exists an explicit 1-soliton solution,
\begin{equation}
W= \sqrt[4]{\frac{\theta_V}{2\theta_S}}~(x+\mathrm{i}y),
\end{equation}
where the $W$ field is the stereographic projection of $\ph$ on the
complex plane, given by $W=2(\phi_1+\mathrm{i}\phi_2){(1-\phi_3)}^{-1}$.
We choose $\theta_S=\theta_V=\frac{1}{2}$ where its total energy equals
$4\pi(1+\frac{8}{3\sqrt{2}})\approx 36.2618$. Since the soliton profile
has a polynomial decay, we need a large lattice. With a $350\times350$
grid and lattice spacing $h=0.05$, we obtain $E=36.4890$ and
$B=0.9999$. Here, the energy is slightly higher than the exact
solution because of the finite lattice effects. The holomorphic baby
skyrmion has the slowest decay of any of the models discussed, and
therefore can be seen as the worst case scenario.

\begin{table}
\begin{center}
\begin{tabular}{|c||c|c|c||c|c|c|}
\hline & \multicolumn{3}{|c||}{1-vacuum model} & \multicolumn{3}{|c|}{2-vacua model} \\ \cline{2-7}
Topological & \multicolumn{2}{|c|}{SA} & Ref.~\cite{baby_pots} & \multicolumn{2}{|c|}{SA} & Ref.~\cite{baby_pots} \\
Charge & $B$ & $E/B$ & $E/B$ & $B$ & $E/B$ & $E/B$ \\ \hline
1 & 0.99978 & 19.6505 & 19.47 & 0.99989 & 19.6572 & 19.65 \\
2 & 1.99973 & 18.4452 & 18.27 & 1.99984 & 17.6530 & 17.65 \\
3 & 2.99962 & 18.5257 & 18.34 & 2.99983 & 17.2259 & 17.22 \\
4 & 3.99952 & 18.4014 & 18.22 & 3.99989 & 17.0677 & 17.07 \\ \hline
\end{tabular}
\caption{Baby Skyrme models: Comparison of our SA results with
Euler-Lagrange results.}
\label{table2d}
\end{center}
\end{table}

\begin{figure}
\begin{center}
\includegraphics[width=11cm]{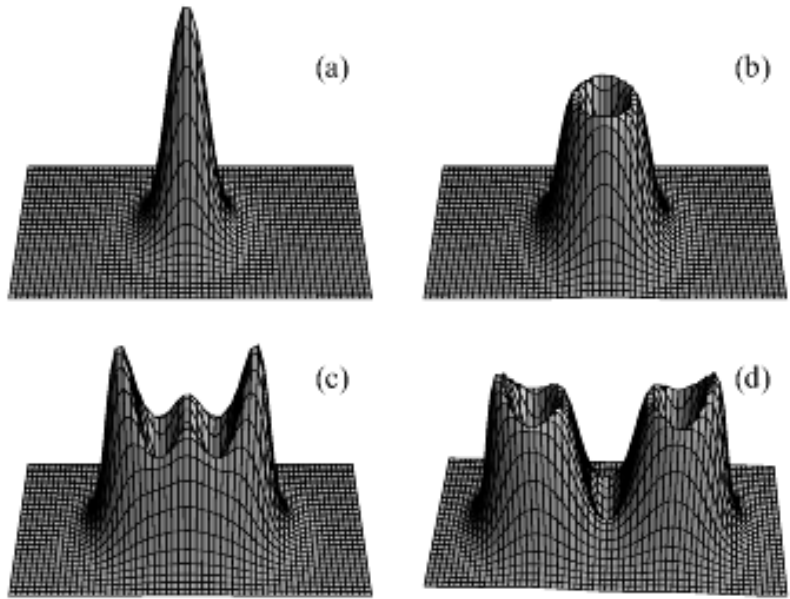}
\caption{Energy density plot with identical vertical energy scale. We
show the 1-vacuum skyrmions of charge one (a), two (b), three (c) and
four (d). The range plotted is $2.4\times 1.2$ units in each figure.}
\label{Baby1-4}
\end{center}
\end{figure}

\begin{figure}
\begin{center}
\includegraphics[width=11cm]{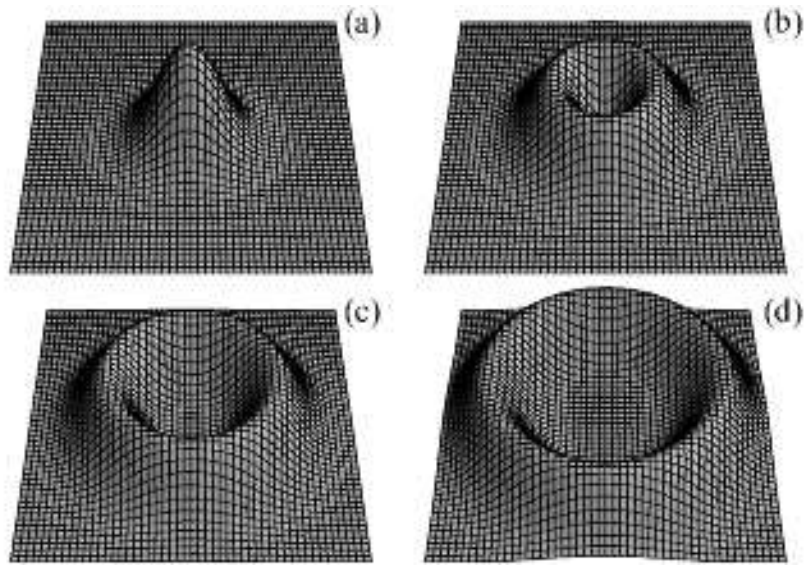}
\caption{Energy density plot with identical vertical energy scale. We
show the 2-vacua skyrmions of charge one (a), two (b), three (c) and
four (d). The range plotted is $2.0\times 1.0$ units in each figure.}
\label{NewBaby1-4}
\end{center}
\end{figure}

We have also looked at the baby Skyrme models with one vacuum, where
$V=1+\phi_3$, and two vacua, where $V=1-\phi_3{}^2$. The parameters for
the 1-vacuum model have been fixed to $\theta_S=\frac{1}{4}$ and
$\theta_V=0.1$ and for the 2-vacua to $\theta_S=0.44365$ and
$\theta_V=0.05$ in agreement with existing literature, see
Ref.~\cite{baby_pots}.  We use an $80\times 80$ grid with periodic
boundary conditions and lattice spacing $h=0.4$.  The minimal energy
solution in the first four topological sectors is shown in
Fig.~\ref{Baby1-4} and Fig.~\ref{NewBaby1-4} respectively.  We compare
the energies per charge with the calculations from
Ref.~\cite{baby_pots} in Table~\ref{table2d}.

\begin{figure}
\begin{center}
\includegraphics[width=16cm]{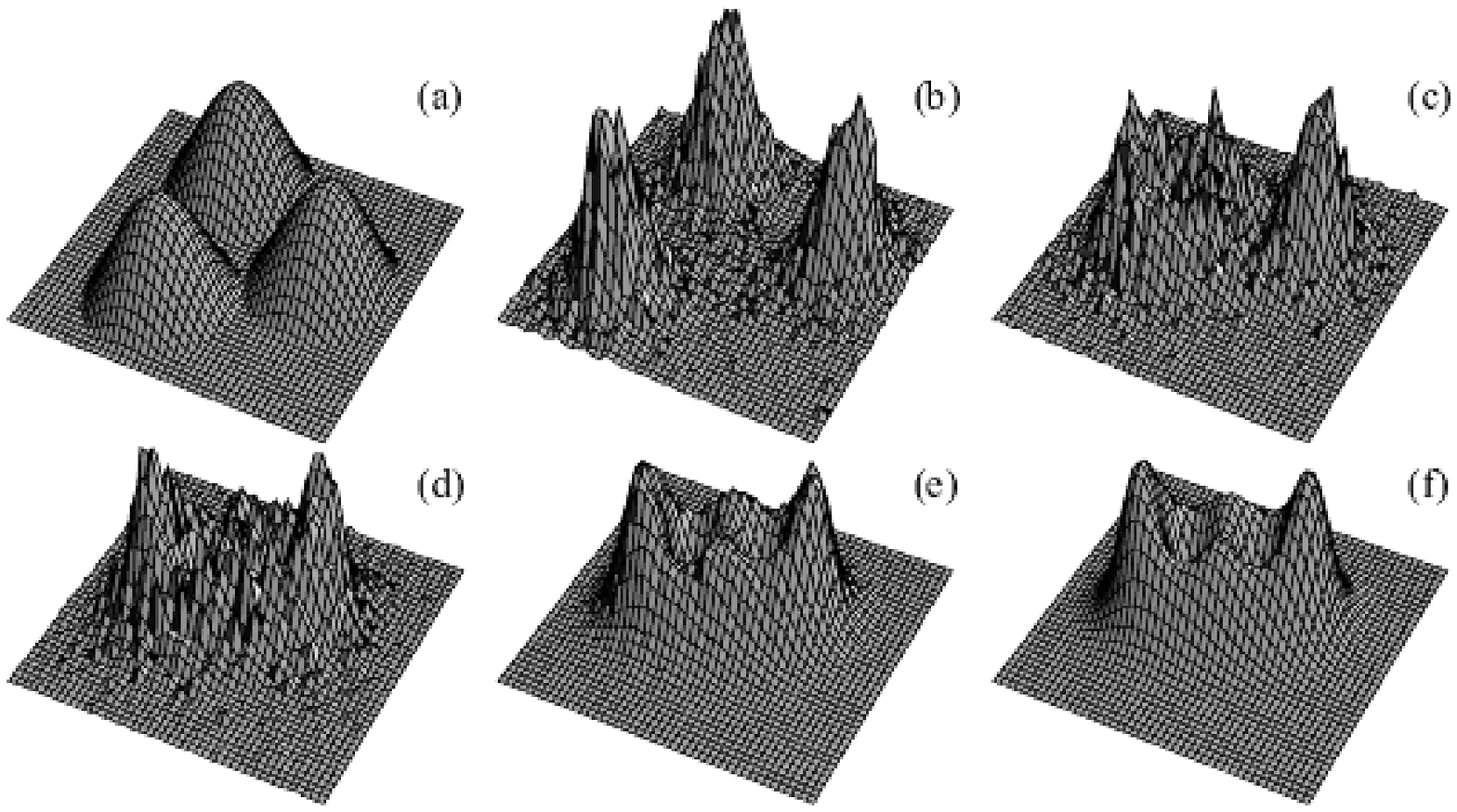}
\caption{Energy density plot of identical vertical energy scale. We
show a SA cooling for the 3-skyrmion in the 1-vacuum model: (a) the
starting configuration; (b) system is heated to $\beta=500$ and
skyrmions repel each other because their isospins are initially in the
same direction; (c) isospins rotate relative to each other and
skyrmions attract each other; (d) equilibrium has been reached for
$\beta=500$; (e) system is cooled to $\beta=5000$; (f) minimal energy
solution at $\beta=\infty$.  The range plotted is $1.8\times 1.6$
units in each figure.}
\label{BCoolSeq}
\end{center}
\end{figure}

The results for the 2-vacua model are the same for both studies within
an accuracy of a few parts in $10^{4}$. The results from
Ref.~\cite{baby_pots} have been obtained via the shooting method. We
can apply this accurate method, because the skyrmions are radially
symmetric and the minimization reduces to a 1D problem.  There is a
slight discrepancy in energy when comparing the 1-vacuum model. The
energies in Ref.~\cite{baby_pots} were obtained on a 2D lattice using
a damped time-evolution.  The energy of the 1-skyrmion calculated
using the shooting method is $E=19.65$. Our SA result agrees well with
this value.  The inaccuracies in Ref.~\cite{baby_pots} arise due to a
different derivative approximation, which tends to reduce the
energy. The errors due to the fixed boundaries are also greater in
Ref.~\cite{baby_pots}, although this tends to increase the energy of
the 1-skyrmion.  We believe that periodic boundary conditions, which
have been used for our SA result, have the advantage that the tails of
the skyrmions can spread out further. The finite lattice effects are
still present, as skyrmions could then interact with themselves over
the boundaries. For our SA model, a study \cite{sc:second} on the
1-skyrmion case shows that the gridsize used induces an error of the
order 0.01\%. The error due to not having relaxed the system
properly is 0.01\%. The largest error is due to finite difference
effects and has a possible error of 0.1\%. Because of the successful
agreement of our SA result for the 1-skyrmion with the result of the
shooting method, it is believed that the SA solutions with higher
topological charge are also more accurate than the results quoted in
Ref.~\cite{baby_pots}. Finally, we show shows an example of the
cooling schedule for its 3-skyrmion in Fig.~\ref{BCoolSeq}.

We studied three different baby Skyrme models. Changing from one
potential to the other could not have been easier. In the case of the
iterative techniques, changing the differential equation is in itself
conceptually easy, but, in practice, a lot of time is spent on getting
the coefficients right and checking the derivation.  For future
research, we intend to use SA to do a systematic check on the
multi-skyrmion structure. Specifically, we are interested in the class
of potentials that leads to radially symmetric multi-skyrmions.

\section{Simulated Annealing in 3D}

We have chosen the nuclear Skyrme model \cite{holzwarth:review} for
our 3D SA implementation, because we believe that SA is a flexible
tool for exploring the multi-skyrmion structure.

In the sixties, Skyrme constructed an effective  field theory of
mesons where the baryons are the topological solitons of the
theory. Research by 't Hooft and Witten  has established that the
nuclear Skyrme model shows important similarities to the low-energy
effective lagrangian of QCD \cite{thooft:2,witten}. The 1-skyrmion can
be interpreted as a nucleon with reasonable success \cite{anw}.  The
numerical work by Braaten, Townend, and Carson \cite{braaten}, and Battye and
Sutcliffe \cite{multi:num} on the structure of classical
multi-skyrmions supports the idea that an  appropriate quantization of
these minimal energy solutions for a given topological sector could
possibly lead to an effective description of atomic nuclei.  However,
the calculation of quantum properties of multi-skyrmions is very
difficult. Part of this is due to the fact that these minimal energy
solutions are not radially symmetric and the theory is
non-renormalisable.  This is rather frustrating, for the claim that
the Skyrme model, descending from a large $N$-QCD approximation models
mesons, baryons and higher nuclei is a very attractive one.  Numerical
methods are probably the only way forward and the SA scheme might be
useful in exploring further the multi-skyrmion structure for different
versions of the Skyrme model.

\subsection{The Nuclear Skyrme Model}

The nuclear Skyrme lagrangian
\begin{equation}
\mathcal{L}=\frac{1}{2}\partial_{\mu}\ph\cdot\partial^{\mu}\ph
-\frac{1}{4}\left[(\partial_{\mu}\vec{\phi} \cdot
\partial^{\mu}\vec{\phi})^{2} -(\partial_{\mu}\vec{\phi} \cdot
\partial_{\nu}\vec{\phi}) (\partial^{\mu}\vec{\phi} \cdot
\partial^{\nu}\vec{\phi})\right]
\label{FullSkLag}
\end{equation}
is a straightforward extension of the nonlinear $\sigma$ model
containing an additional fourth-order term called the Skyrme term. We
need to include this extra term to ensure stability of the
soliton. The mapping becomes
\begin{equation}
 \phi(t): \mathrm{S}^3 \longrightarrow \mathrm{S}^3.
\end{equation}
More realistic lagrangians should probably include higher order
correction terms. The SA scheme is especially useful, because extra
term can be included trivially.

\subsection{Implementation}

The 3D implementation is very similar to the 2D code and we will only
mention new issues relevant to the 3D case.  On a 4D unit sphere, the
integration measure is given by
\begin{eqnarray}
\int_{\rm unit~sphere}\!\d x~\d y~\d z~\d w &=&
\int_{\theta=0}^{2\pi}\int_{\phi=0}^{\pi}\int_{\chi=0}^{\pi}\!
\sin\theta\sin^{2}\chi ~\d\phi~\d\theta~\d\chi\nonumber\\
&=& \int_{\theta=0}^{2\pi}\int_{\cos\theta=-1}^{1}
\int_{\frac{\chi}{2}-\frac{1}{4}\sin(2\chi)=0}^{\frac{\pi}{2}}\!\d\phi
~\d\cos\theta~\d\left(\frac{\chi}{2}-\frac{1}{4}\sin(2\chi)\right)\nonumber\\
&=& \int_{\theta=0}^{2\pi}\int_{u=-1}^{1}\int_{v=0}^{\frac{\pi}{2}}\!\d\theta~\d u~\d v.
\end{eqnarray}
In order to rotate the new vector from the intrinsic frame to the
laboratory frame, the angle $\chi$ must be evaluated from $v$. The
equation
\begin{equation}
 v=\frac{\chi}{2}-\frac{1}{4}\sin(2\chi)
\end{equation}
can not be rewritten in terms of $\chi$, and therefore it must be
solved numerically. The observation that most solutions will be in the
region of small $v$ due to the importance sampling, implies that a
small $\chi$ approximation might be useful. For small $\chi$,
$v\approx\frac{\chi^{3}}{3}$. Therefore, the inversion is performed
numerically by tabulating $\chi$ uniformly on $[0,\pi]$ against
$v^{1/3}=\left(\frac{\chi}{2}-\frac{1}{4}\sin(2\chi)\right)^{\frac{1}{3}}$.
The $\chi$ value corresponding to the selected $v^{\frac{1}{3}}$ is
found in this table, and a linear interpolation is applied between the
two nearest values of $v^{\frac{1}{3}}$ to give a better approximation
to $\chi$. Using 1000 pre-calculated entries the error on the whole
region $\chi\in[0,\pi]$ is then less than $10^{-7}$. On the region
$\chi\in[0,0.1]$, where almost the entire selection of $\chi$ lies,
the error is less than $10^{-14}$, which corresponds to the precision
of double real numbers. This method does not create any significant
errors and is time-consuming. No faster method seems to be possible.

As the cosines and sines of $\phi$, $\theta$, and $\chi$ are known,
the rotation can be performed using a rotation matrix similar to
(\ref{BabyRotationMatrix}). The new choice of vector in the intrinsic
frame is given by
\begin{equation}
 \vec{n}^{\rm int} =(n_{1}^{\rm int},n_{2}^{\rm int},n_{3}^{\rm int},n_{4}^{\rm int})
 =(\sin\chi\sin\theta\cos\phi,\sin\chi\sin\theta\sin\phi,\sin\chi\cos\theta,
 \cos\chi).
\end{equation}
The $z$-axis in the intrinsic frame coincides with $\vec{k}^{\rm
lab}=\ph_{\rm present}(x_k)$ in the laboratory frame, similarly as in
the baby Skyrme model. The rotation angles $\gamma$, $\alpha$, and
$\beta$ (in that order) are defined by
\begin{equation}
 \vec{k}^{\rm lab}=(\cos\beta\sin\alpha\sin\gamma,\sin\beta\sin\alpha\sin\gamma,
\cos\alpha\sin\gamma,\cos\gamma).
\end{equation}

The transformation $\vec{n}^{\rm int}$ to $\vec{n}^{\rm lab}$ is
performed using the matrix
\begin{equation}
\left(\begin{array}{c}
n_{1}^{\rm lab} \\
n_{2}^{\rm lab} \\
n_{3}^{\rm lab} \\
n_{4}^{\rm lab}
\end{array}\right)=
\left(\begin{array}{cccc}
\cos\alpha\cos\beta & -\sin\beta & \sin\alpha\cos\beta\cos\gamma & \sin\alpha\cos\beta\sin\gamma \\
\cos\alpha\sin\beta & \cos\beta & \sin\alpha\sin\beta\cos\gamma & \sin\alpha\sin\beta\sin\gamma \\
-\sin\alpha & 0 & \cos\alpha\cos\gamma & \cos\alpha\sin\gamma \\
0 & 0 & -\sin\gamma & \cos\gamma
\end{array}\right)
\left(\begin{array}{c}
n_{1}^{\rm int} \\
n_{2}^{\rm int} \\
n_{3}^{\rm int} \\
n_{4}^{\rm int}
\end{array}\right).
\label{SkRotationMatrix}
\end{equation}
For similar reasons as already discussed previously, we use a
restricted transition probability to be uniform over $[0,A)$, where
$A\le\pi$. The method of importance sampling has also been
investigated for the nuclear Skyrme model and is given in the
appendix. The transition probability is therefore
\begin{equation}
T(Y\mid X)=\left\{\begin{array}{cl}
\frac{1}{4\pi A} & {\rm if}~\frac{\chi}{2}-\frac{1}{4}\sin(2\chi)<A \\
0 & {\rm otherwise.}
\end{array}\right.
\end{equation}
The angles are sampled by
\begin{eqnarray}
 v&=&\frac{\chi}{2}-\frac{1}{2}\sin(2\chi)=A\xi_{1},\nonumber\\
 u&=&\cos\theta=2\xi_{2}-1,\nonumber\\
 \phi&=&2\pi\xi_{3},
\label{3dsampling}
\end{eqnarray}
where $\xi_{1}$, $\xi_{2}$ and $\xi_{3}$ are three random variables
uniformly sampled from 0 to 1.  The new vector
$\ph_n(x_k)=\vec{n}^{\rm lab}$ is a vector which has been selected
from a uniform probability distribution centered around the previous
vector $\ph_p(x_k)~(=\vec{k}^{\rm lab})$ where only the angle $\chi$
between these vectors has an upper limit. Finally, $\ph_n(x_k)$ and
$\ph_p(x_k)$ are inserted into (\ref{Zaccept}) to find the acceptance
rate.  Similarly to the baby Skyrme models, the value of $A$ is
automatically chosen to have an acceptance rate near 40\%. The cooling
is also controlled in the same manner as described in 2D.

\subsection{Results and Comparison}

\begin{figure}
\begin{center}
\includegraphics[width=16cm]{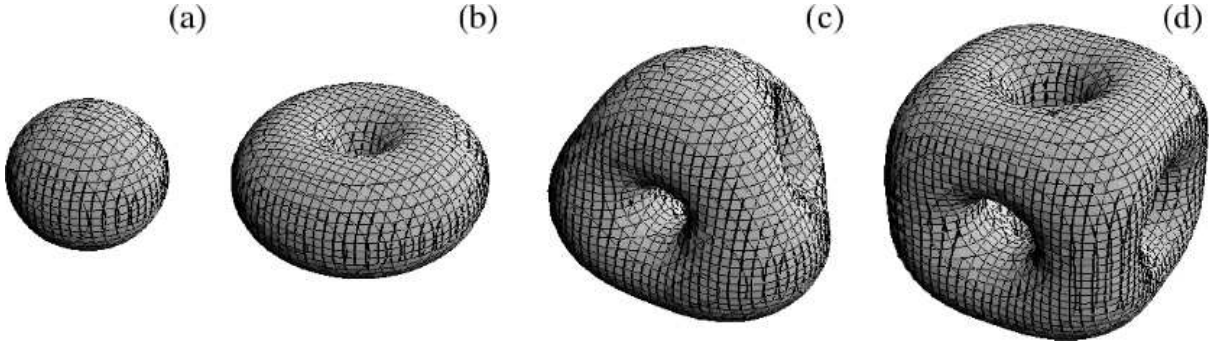}
\caption{Plot of the same constant energy density surface. We show the
first four multi-skyrmion solutions (from left to right). All plots
are to the same scale. The mesh spacing of the plotted objects is
0.12 units.}
\label{FullSk1-4}
\end{center}
\end{figure}
\begin{figure}
\begin{center}
\includegraphics[width=15cm]{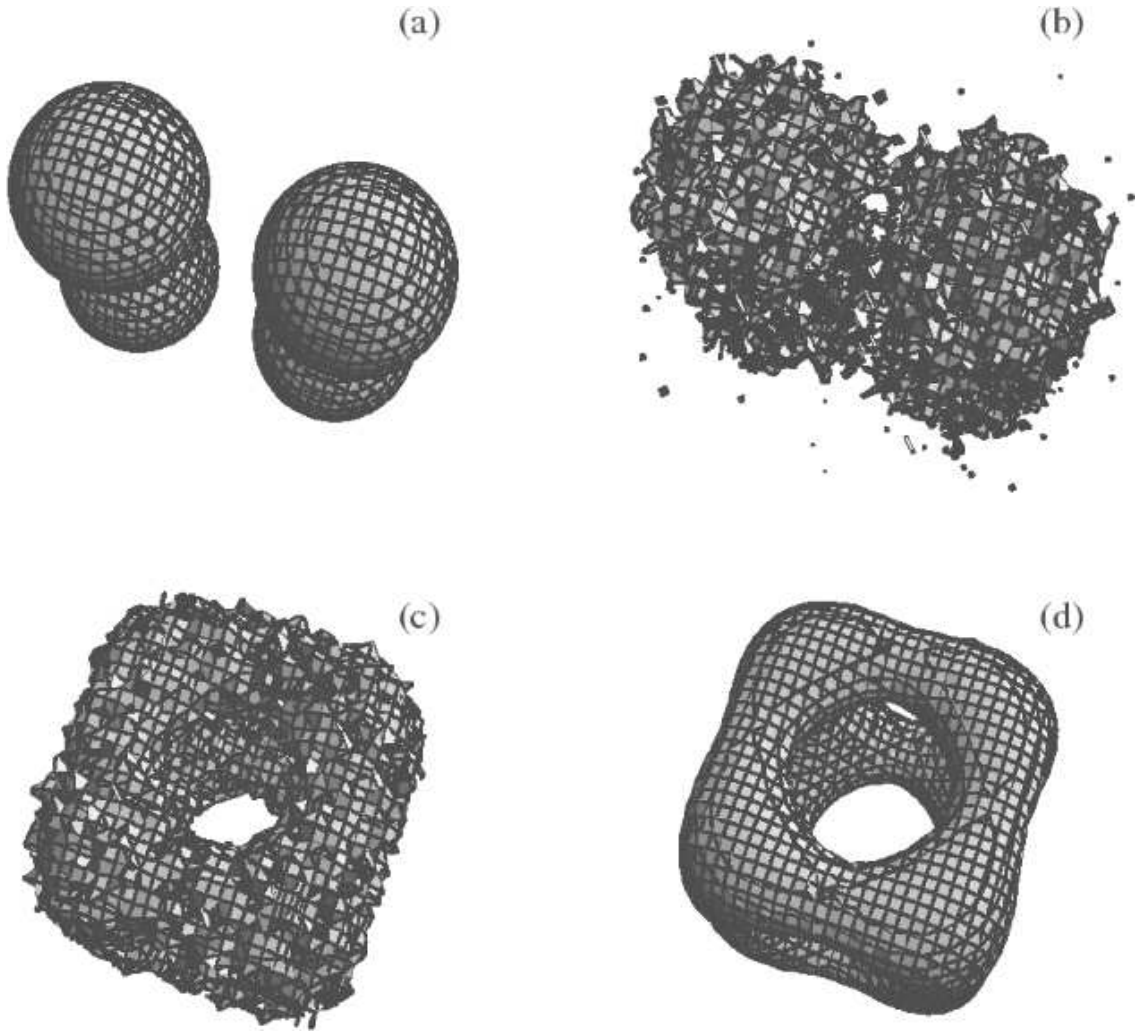}
\caption{Plot of constant energy density surface. We show a SA search
of the $B=4$ skyrmion: (a) the starting configuration of four
1-skyrmions; (b) the system heated to $\beta=500$ where the skyrmions
fuse into one; (c) the system in equilibrium at $\beta=500$ where the
structure emerges; (d) the minimal energy solution at $\beta=\infty$.
All plots are to the same scale. The mesh spacing of the plotted
objects is 0.1 units.}
\label{3Dcooling}
\end{center}
\end{figure}
The 3D implementation of SA is computationally much more intensive
than the 2D case.  The accuracy of our numerical simulations is
therefore reduced due to limited resources in computation time and
memory. The intergrid spacing used is 0.12 Skyrme units and is close
to the upper limit where the numerics break down (at a reasonably high
temperature required to do SA sufficiently fast). The maximum gridsize
that can be used to obtain results in a reasonable time is $80\times
80\times 80$. The finite volume causes an increase of energy for the
1-skyrmion because it repels itself over the periodic boundary, and
therefore induces an error of 1\% \cite{sc:second}. The error due to
not having relaxed the system properly is 0.1\%. The error due to
finite difference effects has a maximum of 0.3\%.
The skyrmions of topological charge one to four are shown in
Fig.~\ref{FullSk1-4}. We show an an example of the cooling schedule
for the 4-skyrmion in Fig.~\ref{3Dcooling}.  These energies per charge
are contrasted in table~\ref{table3d} with those results obtained by
Battye and Sutcliffe \cite{multi:num}. It is very difficult to compare
the results.  The 1-skyrmion solution gives more information for
comparison. It is spherically symmetric and the shooting method in the
hedgehog ansatz can be used.  The energy of a 1-skyrmion minimized in
the 1D SA code is $E=73.12$ (in the continuous limit). We are not sure
if the result for the 1-skyrmion in Ref.~\cite{multi:num} is truly
more accurate or just a coincidence. Their topological charge, an
indicator for the discretization error, is certainly less accurate.
The finite lattice effect increases the energy of the 1-skyrmion and
therefore the energy obtained should be larger than 73.12.
Unfortunately, we do not know which lattice parameters they have used,
making a good comparison impossible. However, we get the same minimal
energy structure.
\begin{table}
\begin{center}
\begin{tabular}{|c||c|c|c|c|} \hline
Topological & \multicolumn{2}{|c|}{SA} & \multicolumn{2}{|c|}{Ref.~\cite{multi:num}} \\
Charge & $B$ & $E/B$ & B & $E/B$ \\ \hline
1 & 1.0015 & 73.75 & 0.984 & 72.96 \\
2 & 2.0030 & 70.31 & 1.972 & 69.34 \\
3 & 3.0042 & 68.52 & 2.960 & 67.69 \\
4 & 4.0048 & 66.30 & 3.948 & 66.09 \\ \hline
\end{tabular}
\caption{The nuclear Skyrme model: SA results versus published results.}
\label{table3d}
\end{center}
\end{table}

\section{Conclusion}

We have shown that SA is an alternative way of finding the minimal
energy solution in a given topological charge sector. We
independently confirmed the validity of the studies using the standard
minimization techniques. It is very hard to objectively compare the
different approaches. However, we have found SA to be a more
convenient and flexible minimization technique. The implementation
and fine-tuning of our SA codes took a fair amount of time due to a
lack of prior research in this area. In comparison to other methods,
we are confident that future implementations will take us
considerably less time. We did not find any significant differences
in speed of minimization. The SA codes can be made faster by
fine-tuning the cooling parameters. We prefer SA minimization because
of its ease of use. Speed considerations are irrelevant in 1D and 2D
and we can use parallel computing in the 3D case.
There are several areas we want to look at next. First of all, we will
optimize SA by using more sophisticated update and cooling mechanisms
and by parallelization. We are also currently investigating the
possibility of doing time-evolution via SA minimization of the action.
At the same time, we intend to look at a wide range of models.  We
shall investigate the multi-skyrmion structure of several baby Skyrme
models. Research is also underway in the use of symmetry
breaking terms for the nuclear Skyrme model. Moreover, the 2D and 3D
code will be used to study phase transitions in the baby and nuclear
Skyrme model at finite temperatures and finite densities
\cite{os:finite}. To conclude, SA is a flexible tool; all we really
need is an energy functional to minimize!

\section*{Acknowledgements}

First of all, the authors would like to thank Niels Walet for his
support and advice. We would also like to acknowledge discussions with
Klaus Gernoth, Laur J\"arv and Bernard Piette. The authors acknowledge
PPARC (M.~H.) and EPSRC (O.~S.~\& T.~W.) for research grants.

\appendix
\subsection*{Appendix: Importance Sampling}

For importance sampling, the integral (\ref{average}) is rewritten as
\begin{equation}
 \langle\mathcal{F}\rangle=\int\left[\frac{\mathcal{F}(C)P(C)}{\tilde{P}(C)}\right]\tilde{P}(C)\,\d C.
\end{equation}
Here $P(C)$ can be normalized without loss of generalization, see
Ref.~\cite{ka:monte}, and $\tilde{P}(C)$ is a different probability
density function which satisfies
\begin{equation}
 \tilde{P}(C)\ge 0,~~~~~\int\tilde{P}(C)\,\d C=1,
\label{constrftilde}
\end{equation}
and
\begin{equation}
 \frac{\mathcal{F}(C)P(C)}{\tilde{P}(C)}<\infty
\end{equation}
except on a countable set of points. In this method  one chooses a
$\tilde{P}(C)$ that minimizes the variance which is
\begin{equation}
{\rm var}\{\langle\mathcal{F}\rangle\}=\int\frac{\mathcal{F}^{2}(C)P^{2}(C)}
{\tilde{P}^{2}(C)}\d C-{\langle\mathcal{F}\rangle}^{2}.
\end{equation}
The measurement of statistical accuracy is given by
${\rm var}\{\langle\mathcal{F}\rangle\}$.
More samples reduce the variance. Alternatively, the same
variance using fewer samples can be achieved with importance
sampling. In practice, the closer $\tilde{P}(C)$ is  to $\mathcal{F}(C)P(C)$,
the lesser the variance becomes.  It is known that if
\begin{equation}
 \tilde{P}(C)=\frac{\mathcal{F}(C)P(C)}{\langle\mathcal{F}\rangle},
\label{ftildemin}
\end{equation}
then the integral is equal to $\mathcal{F}$ with zero variance. However,
we need to respect the constraints (\ref{constrftilde}), and even
worse, choosing a $\tilde{P}$ requires knowledge of $\langle\mathcal{F}\rangle$ prior
to evaluating the integral.

\subsubsection*{Baby Skyrme models}

The application of importance sampling to the partition function
(\ref{Zfunction}) is complicated, because the range of integration is
on a sphere of unit length. Therefore, we need to use a
$\tilde{P}(\ph(x_k))$ that is only non-zero  on the unit
sphere. Further, the maximum or most likely area of accepted values
depends on the present vector, and therefore $\tilde{P}(\ph_n(x_k))$
should not be restricted to a certain region of the sphere, but should
depend on the present vector $\ph_p(x_k)$. Looking at the results of a
uniform probability distribution function on the sphere, a Gaussian
distribution of the polar angle $\theta$ seems to be a good choice for
$\tilde{P}$, where $\theta=0$ is in the direction of the present
vector. This distribution is then rotated around the azimuths
$\phi$-axis and therefore the $\phi$-distribution is uniform. The
Gaussian-distributed $\theta$ and the uniformly distributed $\ph$
define the probability for the new vector. This vector is inserted
into $\mathcal{F}(C)$ as before, and the Metropolis algorithm accepts or
rejects this particular choice. The quantity $q$ is given by
\begin{equation}
q(C_2\mid C_1)=\frac{P(C_2)\tilde{P}(C_1)}{P(C_1)\tilde{P}(C_2)}.
\label{qImpSamp}
\end{equation}
To implement this method, a Gaussian is chosen centered along the $z$-axis in
the intrinsic frame. This Gaussian is, in terms of the $z$-coordinate,
\begin{equation}
 \tilde{f}=\exp[-A(1-\cos\theta)^{2}]=\exp[-A(1-z)^{2}].
\label{tildef}
\end{equation}
The integral
\begin{equation}
 \int_{-1}^{1}N\exp[-A(1-z)^{2}]\,\d z
\end{equation}
satisfies (\ref{constrftilde}) where $N$ is a normalization constant
and $A$ is an arbitrary parameter that changes the breadth of the
Gaussian and thereby alters the acceptance rate. This is sampled using
the Box-M\"{u}ller method, by choosing
\begin{equation}
 z=1-\frac{1}{\sqrt{A}}\sqrt{-\ln\xi_{1}}|\cos(2\pi\xi_{2})|,
\end{equation}
which is a shifted Gaussian so that the peak is at
$z=\cos\theta=1$. All values of $z<-1$ are rejected. The azimuth angle
$\phi$ is sampled uniformly along $[0,2\pi)$ by $\phi=2\pi\xi_{3}$.
The acceptance probability becomes
\begin{equation}
A(C_2\mid C_1)=\min\left(1,\exp \left[-\beta(\mathcal{V}_{n}-\mathcal{V}_{p})
+A(1-z)^{2}\right]\right),
\end{equation}
using (\ref{qImpSamp}) and (\ref{tildef}).

\subsubsection*{Nuclear Skyrme model}

We do importance sampling by prioritizing small $v$ values and
selecting $u$ and $\phi$ with uniform probability. The small $v$
region corresponds to the small $\chi$ region. Importance sampling is
used because the newly selected four dimensional unit vector should be
in the neighborhood of the previous vector. This new vector is
selected in the intrinsic frame, as discussed in Sec.~\ref{MCBabyImportance},
where the previous vector is pointing in the
$\chi=0$ direction. A good probability distribution is
\begin{equation}
 \tilde{f}=\e^{-A v}.
\label{Sktildef}
\end{equation}
The quantity $v$ is therefore selected using
\begin{equation}
 v=-\frac{1}{A}\log\left(\left[\exp\left(-\frac{\pi}{2}A\right)-1\right]\xi_{1}+1\right),
\end{equation}
where $\xi$ is a uniform random variable on $(0,1)$. The other angles
are sampled as in Eq.~(\ref{3dsampling}).
The acceptance probability now becomes
\begin{equation}
A(C_2\mid C_1)=\min\left(1,\exp\left[-\beta(\mathcal{V}_{n}-\mathcal{V}_{p})+A v\right] \right).
\end{equation}
A similar probability distribution function can be used for the baby
Skyrme model and is faster than the given gaussian.  In practice
importance sampling is not used, because it is computationally more
time-consuming than restricting the transition probability.

\bibliographystyle{prsty}
\bibliography{references}
\end{document}